# A Survey on $4\pi$ Treatment Planning Technique for Radiation Therapy

Amir Moslemi[a,b], Navid Khaledi[a]

[a]Department of Physics, Ryerson University, Toronto, Canada

[b]Physical Sciences, Sunnybrook Health Sciences Centre, Toronto, M4N 3M5, Ontario, Canada

**Corresponding Author:**

Amir Moslemi, PhD
Physical Sciences, Sunnybrook Health Sciences Centre
2075 Bayview Ave, Toronto, Ontario M4N 3M5
Email: Amir.moslemi@ryerson.ca

## Abstract

The challenge of removing cancerous cells lies in the limitation of organ at risk, which restricts the ability to increase the radiation dose adequately for enhancing treatment effectiveness. This survey provides a comprehensive overview of the 4π planning technique for radiation therapy. In radiation therapy, the gantry of Medical linear accelerators can rotate around the couch to administer radiation from various angles. Researchers have introduced the 4π technique with the objective of increasing the radiation dose to tumors while minimizing damage to nearby organs. While dose escalation may not be feasible in certain cases due to potential complications in normal tissues, the 4π technique enables dose escalation for certain types of cancers. The optimization problem formulation constitutes a major innovation in the 4π technique. In this survey, we went thorough optimization problem of 4π technique and discussed about the available investigations that have been performed on 4π technique.

# Introduction

Using a number of fixed radiation beams on different planes for tumor treatment can be effective in terms of organ at risk dose reduction. To this end, the concept of non-coplanar radiation therapy, with aim of using radiation fields without overlapping geometric planes, was introduced [1]. Non-coplanar radiation therapy has been common in stereotactic body radiation therapy and single fraction radiosurgery. The main objective of this technique is to deliver the maximum dose to area of planning target volume (PTV) and sharp dose reduction outside of PTV. Non-coplanar beams were utilized for in accelerated
partial breast irradiation (APBI) [2], and head and neck cancers [3] to increase performance of treatment and minimize the complications for normal tissues. This should be noted that Non-coplanar techniques are being used for prostate cancer and pancreatic cancer treatment. Table 1 illustrates abbreviations which are used in this survey.

Table 1. List of abbreviations which are utilized throughout of this survey.

| Abbreviation | Complete Form |
| --- | --- |
| GTV | Gross tumor volume |
| CTV | Clinic tumor volume |
| PTV | Planning target volume |
| OAR | Organ at risk |
| IMRT | Intensity modulation radiation therapy |
| VMAT | Volumetric modulation arc therapy |
| SBRT | Stereotactic body radiation therapy |

In terms of biological considerations, local tumor controlling requires high biological equivalent doses (BED). Conventional fractionated radiotherapy regimens have relatively low tumor control due to low BED. To tackle this challenge, SBRT was used to deliver a higher BED to tumor with aim of tumor local controlling using a smaller number of high dose treatment fractions. A subtle point should be noted is the correlation between overall survival of patient and local control with SBRT[4]. Additionally, studies show that SBRT is a promising technique in treatment of cancers such as hepatocellular carcinoma (HCC) treatment, specifically for patients with unresectable primary HCC [5,6]. However, organs at risk (OARs) are important limitation in delivering high dose to tumor. Since tumors are surrounded by normal tissues, dose delivery must be precisely accomplished. Although radiation can be a cause of complication for liver, liver SBRT can be successful by conforming the dose to clinical tumor volume (CTV). As a result, clinical trial groups recommended to use greater number of beams to deliver the dose from different directions in order to achieve dose conformality. Due to the overlap of enter or exit doses of different beams in coplanar radiotherapy, this method is not the most effective approach in sparing OARs[7,8]. For first time in 2008, non-coplanar beams were selected to enhance the plane dose conformality. Non-coplanar technique with a large number of filed is not manually possible due to huge search space. In huge search space, obtaining optimum solution can be difficult and it is costly because of the high computation time. Then, computation complexity and order of complexity are the most important to find optimum solution and in worse case scenario, beams selection in huge search space can considered as non-deterministic polynomial times problem which is called NP-hard problem in optimization. Beam orientation optimization (BOO) and fluence map optimization (FMO) are two main terms of radiation therapy treatment planning optimization problem. BOO technique is leveraged to specify and penalize the overlapping between OAR and beams. Filtered back projection can be used for BOO aim [9]. On the other hand, FMO was separated from BOO to decrease complexity of problem [10]. Although utilizing of geometric technique is simple and fast, achieving global optimum solution is not possible due to BOO-FMO decoupling. In other words, the search space is not convex due to BOO-FMO decoupling and getting stuck in local optimum is the main and big challenge. Since there is an intrinsic difference between beam-organ overlapping and organ dose, an objective function based on organ-dose, an objective function based on organ-dose is constructed to improve combination of BOO and FMO. Consequently, the objective is to obtain the optimal non-coplanar (NC) planes with integration BOO and FMO. To circumvent this challenge, Dong et all proposed $4\pi$ NC[11].

Then, a study was launched and led by a group in UCLA for $4\pi$ NC IMRT [11-12].

### $4\pi$ NC IMRT

For the specific case of non-coplanar radiotherapy, the solid angle of treatment using C-arm linear accelerators (Linac) which includes head of Linac rotation, and treatment couch movement, can be considered equal to $4\pi$. Then, it can be named $4\pi$-NC. Dong et all proposed $4\pi$ NC delivery technique for liver SBRT. There are 1162 NC isocenter beams with 6º separation angle between two adjacent beams in a 4π solid angle space [11].

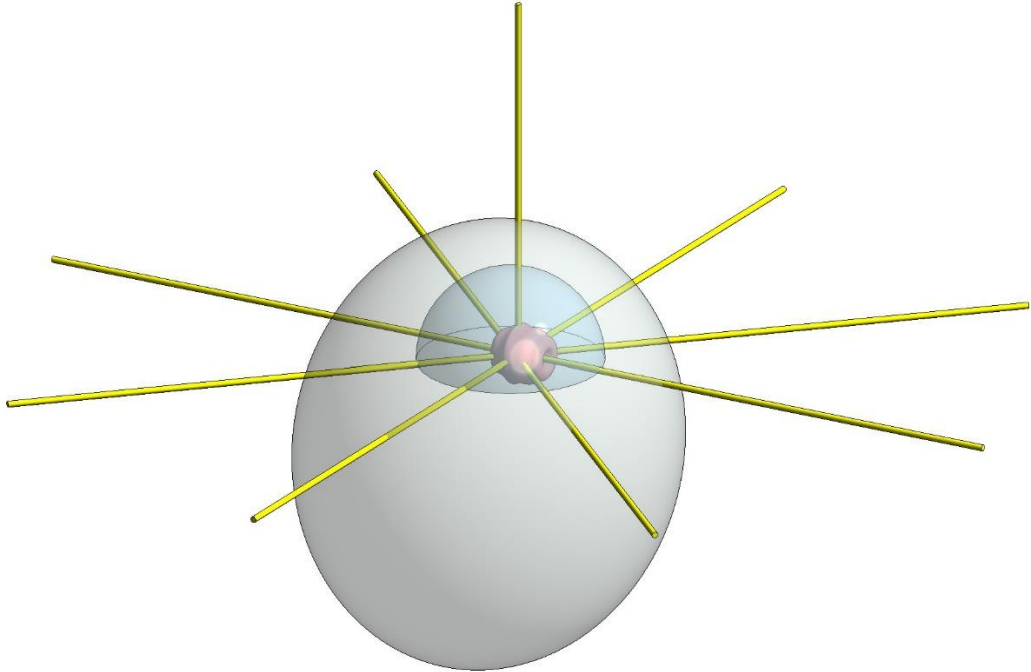

Figure 1. Big grey sphere, light blue half-sphere, pink shape, and yellow lines are head, brain, tumor or PTV and beams, respectively. yellow lines show the selected beams for treatment.

Angle between adjacent beams indicates resolution. A study showed that finding angular resolution does not improve dose distribution [13]. All the beams, which could cause collision between the gantry and couch or patient, are eliminated using computer assisted design (CAD) simulation. CAD was utilized to digitize the treatment room and patient geometry. The geometry of 4π beams is illustrated in Figure 1 and it is sketched by auto-desk Revit 2020().

### 4π Optimization Objective Function

The formulation of IMRT optimization objective function was constructed by Romeijn et al [14].

$$Min\ F(Z)$$

$$s.t$$

$$\text{dose for voxel } j: z_j = \sum_{b \in B} \sum_{i \in B_b} D_{b_{ij}} x_{b_i}$$

$$\text{bixel intensity}: x_{b_i} \geq 0 \quad b \in B', i \in N_b \tag{1}$$

$$x_{b_i} = 0 \quad b \in \frac{B'}{B}, i \in N_b$$

$$z_j \leq q_j \quad q_j \text{ is the dose constraint for voxel } j$$

In this optimization model equation $D_{b_{ij}}$ indicates the dose delivered to a voxel "j" form beam "$i \in N_b$" in beam "$b \in B$".

A greedy algorithm was leveraged to solve (1) to obtain optimum $B'$ such that $B'$ represents selected beam orientation. In this technique, an empty set is taken as solution and in each iteration a new beam from the conformal beam $\frac{B'}{B}$ is added to select beam set as BOO in each iteration and then FMO can be solved. Stopping Criteria to finish iterative process is a challenge for algorithm. For this greedy algorithm, the number of beam or objective function plateau can be considered as iteration stop criteria.

Solving (1) using iterative algorithm is not clinically practical due to heavy computation time. To tackle this challenge, Karush Kuhn Tucker (K.K.T) condition was applied.

$$G_s(Z) = \gamma_s \, mean(Z_j) + (1 - \gamma_s) \max(Z_j) \quad for \quad OARs \tag{2}$$

$$G_r(Z) = \gamma_r \, mean(Z_j) + (1 - \gamma_s) \min(Z_j) \quad for \quad PTVr \tag{3}$$

$$F(z) = \sum_{m \in s,r} \alpha_m G_m(Z) \tag{4}$$

Where $\gamma_s, \gamma_r \leq 0$ and $\alpha_m \geq 0$ for OAR and $\alpha_m \leq 0$ for PTV. There is a trade-off between the plan quality and complexity. In this study, $R_{50}$ (the depth at which the depth dose is 50% of tits maximum value) was used to quantify the dose gradient outside the PTV [14]. Ten liver SBRT patients were re-planned by $4\pi$ NC such that they were previously treated by VMAT. Results of this study significant reduction in normal organ dose was observed, OAR sparing was improved and $R_{50}$ was reduced. Doses to OAR such as left kidney and right kidney and maximum doses to the stomach and spinal cord were considerably reduced. Therefore, this study found that local tumor control can be achieved with a more escalated dose without violating normal tissue constraint. Multiple treatment for metastases can be practical due to low dose imposing to the normal liver volume. Additionally, 4π technique had more flexibility to find OAR-sparing beam angles. The last but not least, coplanar plans restricted to only transverse plane, but non-coplanar utilized all variable beams. Results of this study indicated that normal tissue sparing and dose conformality can be improved by 4π NC IMRT technique compared with VMAT.

As a limitation of this technique, the big challenge of 4π NC is integration of BOO and FMO which is a combinatorial problem. For example, the number of solutions for selecting 14 beams out of 1000 beams is approximately $10^{30}$ which shows the difficulty of this optimization problem.

A comparison study was done to evaluate viability of NC-VAMT for liver SBRT and coplanar VMAT and 4π NC IMRT [15]. In this study, coplanar VMAT, NC-VMAT and 4π IMRT were studied in terms of dose escalation and normal tissue toxicity for 20 patients. The conformity number, homogeneity index, 50% dose spillage volume, normal liver volume receiving greater than 15 Gy, dose to OARs and tumor control probability were considered as criteria to compare the techniques. As mentioned before, tumor locally controlling cannot be achieved by conventional radiotherapy method, but SBRT can be applied to deliver high-dose to tumor in fewer fractions. Coupling VMAT with SBRT can be effective method for liver to deliver high-dose to tumor and spare normal tissue [16]. On the other hand, studied showed that normal tissue sparing is improved by NC beams [17]. Although for NC-VMAT non-coplanar planes were manually selected with the aim of normal tissue sparing and conformity improvement, clinically improvement such as normal tissue complication reduction never observed.

4π radiation therapy was leveraged for automatic NC beam orientation selection and the actual usable beam angles were less than 4π due to collision between patient and couch. Each technique was applied in this study as follows:

Coplanar VMAT: In this technique two full coplanar arcs with 90º collimator angle offset were used.

NC-VMAT: In this technique 3 to 4 partial noncoplanar arc were manually selected.

4π static plans: The first step is to eliminate beams which cause collision between gantry and couch or patient. A trade-off must be considered between plan deliverability and plan quality. In this study, kidney, stomach and spinal cord were considered as OARs for liver. Conformity index as evaluation metric of PTV dose coverage by a certain isodose is defined as follows:

$$CN = \frac{TV_{RI}}{TV} \times \frac{TV_{RI}}{V_{RI}} \tag{5}$$

Where $TV_{RI}$, $TV$ and $V_{RI}$ are target volume covered by the prescription isodose, target volume and volume of the prescription isodose, respectively. $R_{50}$ is defined as follows:

$$R_{50} = \frac{V_{50\%}}{PTV} \tag{6}$$

Where $V_{50\%}$ is 50% of isodose volume. The homogeneity index (HI) was defined as follows:

$$HI = \frac{1+(D_{2\%}-D_{98\%})}{Prescription\ dose} \tag{7}$$

Results of this study showed that the $4\pi$ technique had a better performance than coplanar VMAT and NC-VMAT. Table 2 and table 3 show OAR sparing and evaluation metrics for coplanar VMAT (C-VMAT), NC-VMAT and $4\pi$. These tables iwas provided directly from related research which were reported to provide good information to distinguish different techniques of planning with aim maximum OAR sparing [15].

Table 2. Average OARs doses of 20 patients with coplanar VMAT, NC-VMAT and $4\pi$ plans [15].

|  | OAR Doses (Gy) | | | | | |
| --- | --- | --- | --- | --- | --- | --- |
|  | L-Kidney (mean) | R-Kidney (mean) | Normal Liver (mean) | Stomach (mean) | Spinal Cord (mean) | Body (mean) |
| C-VMAT | 1.42 (2.4) | 2.04 (2.4) | 7.07 (2.6) | 11.15 (6.7) | 6.69 (3.2) | 1.51 (0.6) |
| NC-VMAT | 1.46 (2.1) | 2.09 (2.5) | 6.97 (2.8) | 10.95 (10) | 5.74 (4) | 1.51 (0.6) |
| $4\pi$ | 0.82 (1.2) | 1.7 (1.4) | 6.01 (2.3) | 7.58 (2.3) | 4 (3.7) | 1.46 (0.6) |

L: left, R: right

Table3. Average dosimetric metrics for 20 patients with coplanar VMAT, NC-VMAT and $4\pi$ plans [15].

|  | PTV $D_{98\%}$ (Gy) | $VL_{>15}$ (cm$^3$) | $V_{50}$ (cm$^3$) | $R_{50}$ | HI | CN |
|---|---|---|---|---|---|---|
|  | Evaluation | Metrics |  |  |  |  |
| C-VMAT | 49.62 (9.6) | 233.7 (120.8) | 271.5 (185.6) | 3.66 (0.4) | 0.11 (0.03) | 0.94 (0.03) |
| NC-VMAT | 49.59 (9.7) | 228.7 (122.3) | 287.5 (213.4) | 3.77 (0.5) | 0.1 (0.02) | 0.93 (0.04) |
| $4\pi$ | 49.42 (9.7) | 153.9 (84.5) | 210.8 (146.4) | 2.82 (0.3) | 0.11 (0.03) | 0.93 (0.03) |

$VL_{>15}$ :The normal liver volume receiving a dose higher than 15Gy.

Based on the information of Table 2 and Table 3, the $4\pi$ technique had best performance compared with C-VMAT and NC-VMAT in terms of OAR sparing for all organs at risk. In terms of evaluation metrics, $4\pi$ had a better performance for PTV $D_{98\%}$, $VL_{>15}$ (cm$^3$), $V_{50}$ (cm$^3$) and $R_{50}$ metrics. Superiority of $4\pi$ over C-VMAT and NC-VMAT in terms of 50% dose line was significant. The results of this study proved that dose can be escalated using $4\pi$ more than C-VMAT and NC-VMAT.

In the light of $4\pi$ technique, Deng et al leveraged $4\pi$-NC SBRT for central large lung tumors treatment [18]. As we know, SBRT technique for lung tumors can deliver 50-60 Gy in 3-5 fractions that is why it is called hypo-fractionated doses. The important challenge is that there is a direct correlation between dose increasing and OAR complication probability. In this study, the number of beams selected was selected based on a comparison between coplanar and non-coplanar plans with respect to $R_{50}$. Study was constructed based on a comparison between IMRT, VMAT and $4\pi$-NC technique for 12 patients with lung tumors. Results of this study showed that $4\pi$-NC was significantly better than other planning techniques in terms of maximum dose delivery. In terms of OARs sparing, doses to sensitive organs such as heart and spinal cord were reduced by 32% and 53%, respectively. It means that doses to spinal cord as one of the most important radiosensitive organs has been approximately half in comparison with IMART and VMAT. Consequently, lung SBRT can be improved in terms of dosimetric aspects using $4\pi$-NC planning, and doses to PTV can be considerably escalated. As a result, $4\pi$-NC provided an optimum treatment for central and large lung tumors such that doses to tumors can be increased and doses to OARs can be reduced.

In another study, Murzin et al utilized 4π-NC technique for brain radiotherapy with aim of the cortical sparing [19]. Point should be noted is that the cognitive ability is declined by irradiating of normal sensitive brain tissue. To this end, this study compared 4π-NC planning and IMRT in terms of brain sparing. Thirteen patients with high-grade glioma were investigated. These patients were previously treated by IMRT and 4π-NC optimization was then applied to replan treatment. 4π-NC technique was applied to minimize the cortical dose the PTV and to consider constraint for normal cortex, respectively. 4π-NC optimization in this study was same as the references ([11-12]) such that 1162 beams in 4π angle with 6º separation to each other were considered and optimum beams could be obtained by solving Eq(4). Maximum point dose, mean dose and equivalent uniform doses (EUD) metrics were considered to evaluated treatment planning techniques. The PTV dose conformity index (CI) and homogeneity index (HI) were considered to evaluate dose outside PTV and hot/cold spots inside PTV. CI and HI can be expressed as follows:

$$CI = \frac{V_P}{V_{PTV}} \qquad (8)$$

$$HI = \frac{D_2 - D_{98}}{D_P} \qquad (9)$$

Where, $V_P$ and $V_{PTV}$ are the volume of PTV that is covered by prescribed dose isodose line and PTV volume, respectively. $D_2$, $D_{98}$ and $D_P$ are minimum doses for 2% of PTV, minimum doses for 98% of PTV and prescribed dose, respectively. Results indicated overall doses to all brain OARs are reduced using 4π-NC planning. Doses to cortical and white matter were considerably decreased using 4π-NC in compared IMRT. As a result, 4π-NC planning had significantly better performance than IMRT with aim of OAR of brain dose decreasing. This study was aimed for cortical sparing and results proved that 4π-NC planning can spare cortical such that doses to hippocampus and white matter were considerably reduced. Additionally, 4π-NC planning minimized dose-spillage outside the PTV. Therefore, doses to PTV for brain can be escalated without considerable impact on OAR using 4π-NC planning.

Tran et al made a comparative study among intensity-modulated proton therapy (IMPT), 4π-NC IMRT and VMAT for prostate cancer treatment [19]. In this study not only planning techniques were compared but also a comparison was investigated between photon beam and proton beam. Since the prostate cancer is second leading cause of cancer death for men in the USA, treatment with higher fractionated treatments can be effective method with OAR sparing consideration. IMPT is a technique which leverages pencil beam scanning to optimize all beams based on inverse problem theory for uniformly dose delivery. In terms of $4\pi$-NC technique, 1162 non-coplanar beams such that $6°$ apart between beams was considered. Consequently, 30 beams as optimum NC beams were selected by solving $4\pi$-NC optimization problem. Results of this comparison study showed that all these planning techniques have similar performance in terms of OAR sparing. However, IMPT could deliver high dose to the femoral head. As a result, better performance by IMPT in terms of dose spillage, target dose homogeneity and OAR sparing was reported by this study. Additionally, $4\pi$-NC planning technique was better than VMAT in terms of OAR sparing for all organs at risk except sigmoid colon. This point must be noted that IMPT is more expensive than photon technique.

Using predictive model-based technique can be useful to add informative data to treatment planning system with aim of maximum tolerable dose of organ estimation. To this end, knowledge-based planning (KBP) methods were proposed to utilize treated patients geometric-dosimetric information for new patients' dose prediction [20-22]. Tran et al, in other study, utilized knowledge-based technique for liver SBRT treatment using VMAT and $4\pi$-NC planning technique [23]. In this study 21 patients were investigated such that they were initially treated by VMAT. Then, treatment plans were re-planned by $4\pi$-NC IMRT planning technique to compare VMAT and $4\pi$-NC. The prescribed dose was 30-60 Gy and the liver volumes were between 550-3346$^{cc}$. Patients were initially treated by either 2 coplanar full arcs or 2-3 partial arcs VMAT. Then, treatment plans were repeated by $4\pi$-NC technique. After removing the beams that had collision with gantry, couch or patients using $4\pi$-NC technique, 20 non-coplanar beams were selected out of 1162 NC beams. KBP was utilized to predict liver dose distribution such that dose constraint or a higher bound was considered for dose distribution. Based on clinical data 15 Gy or less (<15Gy) as upper-bound can be considered for liver volume receiving in SBRT. In this study two KBP methods including; overlap volume histogram (OVH) [20] and statistical voxel dose learning (SVDL) [24] were leveraged to extract information about the correlation between patient geometry and dose. Both OVH and SVDL categorized as supervised machine learning technique, and problem can be categorized as regression problem. Result of this study showed that error of prediction for $4\pi$-NC was less than half of VMAT's. The subtle point must be noted is that using optimal non-coplanar geometry for liver SBRT can improve treatment in terms of dose delivery which can be achieved by KBP-4$\pi$-NC.

Isotropic dose distribution is playing significant role for voxel does prediction such that non-isotropic distribution increases the complexity of system and then solving problems for KBP-based methods can be harder, and probability of overfitting can be increased. Although 4$\pi$-NC dose distribution is more isotropic than VMAT, it is not a perfect sphere due to finite number of beams and human body shape. Consequently, non-isotropic aspect for using KBP-methods must be considered.

In another study, Yu et al applied 4$\pi$-NC IMRT technique for treatment of recurrent high-grade glioma patients [25]. Then, 20 beams were selected as optimum number of beams by solving 4$\pi$-NC optimization problem. Brainstem, chiasm, cochlea, eyes, lens and optic nerves were the organs at risk. Results of this study showed that 4$\pi$-NC had a considerable superiority over VMAT in terms of dose spillage in OARs. By comparing mean dose (Gy) of 4$\pi$-NC and VMAT, brainstem mean dose was decreased by 0.94 Gy using 4$\pi$-NC compared with VMAT, and for the rest of OARs the mean dose was reduced using 4$\pi$-NC rather than VMAT. Consequently, results of this study demonstrated that 4$\pi$-NC technique is a valid and optimum technique in terms of dosimetric benefit. Additionally, a safe treatment can be planned by 4$\pi$-NC method such that PTV coverage is not compromised.

Brain tumors treatment is categorized as most important treatment in terms of radiosensitive organs at risk inside brain. If these tumors are recurrent types of tumors, treatment will be more challenging. Recurring glioblastoma multiform (GBM) is categorized as recurrent types of tumors and can be frequently occurred after radiotherapy at the same location. Dose escalation for decreasing GBM recurring has been an important challenge such that doses were increased to 70 -80 Gy without controlling on recurring at the same position [26-27]. Although adding brachytherapy as booster dose in combination with external radiation therapy could deliver nearly 110 Gy total dose, side effects and complications of brain brachytherapy were not negligible and were sever [28-29].

Optical apparatus and brain step are considered as OARs for GBM treatment using radiation. Dose constraints for OARs are defined by RTOG protocol 0825 such that maximum doses of chiasm, lens, brainstem, and optic nerve must be lower than 56, 7, 60 Gy, <55 Gy and mean dose of cochlea must be lower than 45 Gy.

Neguyen et al launched a feasibility study with aim of GBM dose escalation with normal tissue sparing consideration using $4\pi$-NC radiotherapy technique in order to tackle recurring GBM challenge [30]. By solving $4\pi$-NC optimization problem, 30 NC beams were selected for treatment. This study employed three different $4\pi$-NC planning scenarios including; 1) 100 Gy to cover 95% of the PTV ($4\pi PTV_{100Gy}$), 2) 100 Gy to cover 95% of GTV ($4\pi GTV_{100Gy}$), and 3) 5 mm expanding original PTV with the same prescription doses ($4\pi PTV_{PD+5mm}$), to compare $4\pi$-NC with original prescription doses ($4\pi PTV_{PD}$). Results of this study showed that $4\pi PTV_{PD}$ could reduce maximum dose and mean dose of brainstem by 47% and 61%, respectively. where PD is prescription dose. Additionally, $4\pi PTV_{100Gy}$ and $4\pi GTV_{100Gy}$ achieved to a same dose reduction as $4\pi PTV_{PD}$. Whereas, further dose reduction was obtained by $4\pi PTV_{PD+5mm}$ for brainstem such that maximum dose and mean dose reduced by 31% and 52%, respectively. As a result, $4\pi$ plans, except the $4\pi PTV_{100Gy}$, could deliver expected doses to tumor with reducing dose to the brain and brainstem. Established clinical dose for the GBM is nearly 60 Gy and mild dose escalation could not improve local control. However, this study adopted $4\pi$-NC IMRT and could achieve to extreme dose escalation up to 100 Gy. The significant point is that mean and maximum OARs doses were maintained low while 100 Gy delivered to tumor. Therefore, the probability of GBM recurring can be decreased using this technique due to dose to tumor extreme escalation.

### $4\pi$-NC Matrix-Based Formulation Optimization Objective Function

As aforementioned, $4\pi$-NC IMRT treatment planning is categorized as complicated optimization problem that has many hyperparameters. Controlling these hyperparameters increases complexity of this technique and optimum values cannot be obtained by manually. Therefore, KBP can be leveraged to find optimum hyperparameters automatically. KBP-based techniques for radiation therapy treatment planning are similar to conventional machine learning problems such that train-data are geometrical features of prior subjects, and test-data are new patients. Doses for new patients based on their geometrical features can be predicted by KBP which was trained by prior patients. However, availability of sufficient high-quality train-data (plans of prior patients) is a challenge which is directly associated to accuracy of dose prediction. Although $4\pi$-NC planning is an advanced type of IMRT, it needs manually hyperparameters tunning which is a time-consuming action.

To this end, Landers et al proposed an evolve KBP-based $4\pi$-NC IMRT technique to circumvent the manual hyperparameters tuning [31]. Additionally, evolve KBP reduces dependency to high quality plans as train data. As a result, this can be utilized for small training set. In this study, 20 patients with lung cancer and 20 patients with head and neck (HN) cancer were investigated. Evolve KBP is an iterative process such that KBP is applied for dose prediction and each iteration the accuracy of prediction based on plan quality metric (PQM) was evaluated. SVDL technique was also used to predict OARs doses. OAR voxels of training-set were ranked into bins based on Euclidean distance to the PTV and median for each distance bin was considered as new OAR voxels.

In other words, this study formulated evolve KBP-based $4\pi$-NC IMRT as a supervised problem which 3D-dose of voxels are predicted by regression algorithm. In this study, SVDL was compared with spectral regression and support vector machine (SVM). SVDL could achieve to lease mean squared error.

For better understanding of evolve KBP-based $4\pi$-NC planning, manual $4\pi$-NC planning can be formulated as follows:

$$\min_{x} \frac{1}{2}\|l - A_0 x\|_2^2 + \frac{1}{2}\|A_0 x - d\|_2^2 + \sum_{i=1}^{N} \frac{\beta_i}{2}\|A_i x\|_2^2 + \gamma\|Dx\|_1^{(\mu)} + \sum_{b=1}^{B} \omega_b \|x_b\|_2^{1/2} \quad (10)$$

Each term of (10) can be described as follows:

PTV: $\frac{1}{2}\|l - A_0 x\|_2^2 + \frac{1}{2}\|A_0 x - d\|_2^2$

OARs: $\sum_{i=1}^{N} \frac{\beta_i}{2}\|A_i x\|_2^2$

Smoothness: $\gamma\|Dx\|_1^{(\mu)}$

Group sparsity: $\sum_{b=1}^{B} \omega_b \|x_b\|_2^{1/2}$

Where $x_b$ is a vector which carries beamlet intensity value for candidate beam $b$. Vector $x$ is concatenation of the vectors $x_b$ for candidate $B$ beams. As well, $A_0$ and $A_i$ ($i = 1, ..., N$) are dose calculation matrices for the PTV and OARs, respectively. $l$ is the prescription dose vector those elements are associated to PTV voxels, and $d$ vector elements show the maximum dose for each voxel in the PTV.

Moreover, $D$, $\gamma$, $\omega_b$ and $\|.\|_2$ ($\|x\|_2 = (\sum_{i=1}^{n} x_i^2)^{\frac{1}{2}}$) are discrete gradient operator, smoothness weight, group sparsity weight and Euclidean norm (L$_2$-norm), respectively. Additionally, $\|.\|_1^{(\mu)}$ is lasso norm with Huber penalty $\mu$ to control the level of smoothing. Smoothness term was applied to enhance smooth fluence map with aim of deliverability reinforcement. The role of group sparsity term is to enforce the vector $x_b$ elements to zero. In this study, $\omega_b$ was set to select 20 beams [31]. If more than 20 beams were left, k-mean clustering would apply on angels of couch and gantry to determine 20-cluster of beams. Deliverability of 20 NC-beams using $4\pi$-NC technique has clinically been proved [25]. Fast iterative shrinkage-thresholding algorithm (FISTA) was leveraged to solve $4\pi$-NC optimization problem, since objective function (Eq 10) is categorized as non-differentiable problem. FISTA is proposed to tackle challenge of non-differentiable problem [32].

They proposed a method such that hyperparameters are automatically updated that helps avoiding of getting stuck in local optimum of hyperparameters. Consequently, the objective function of automated $4\pi$-NC can be formulated as follows:

$$\min_{x} \frac{1}{2}\|l - A_0 x\|_2^2 + \frac{1}{2}\|A_0 x - 1.05l\|_2^2 + \sum_{i=1}^{N} \frac{1}{2}\|A_i x - \hat{d}_i\|_2^2 + \gamma \|Dx\|_1^{(\mu)} + \sum_{b=1}^{B} \omega_b \|x_b\|_2^{1/2}$$

(11)

This automated $4\pi$-NC objective function restricts PTV voxel dose within 100%-105% of prescription dose $l$. Whereas in manual $4\pi$-NC vector $d$, which was maximum dose should be define by users. On the other hand, OAR voxels were weighted based on hyperparameter $\beta$ in manual $4\pi$-NC, but $\beta = 1$ was considered for automated $4\pi$-NC for all OARs. OAR voxels were also restricted by precited voxel dose $\hat{d}$. As a result, manual setting of $\beta$ parameter was not required.

The proposed method in this study was evolved KBP which means the output updated in each iteration. As a result, a metric must be considered to evaluate results in each iteration. To this end, Nelms et al introduced plan quality metric which was based on a list of dosimetric criteria [33]. A subtle point must be noted is that plan quality metric function is not convex and, consequently, cannot be added to automated 4π-NC planning objective function. Therefor, plan quality metric was only considered at end of each iteration to evaluate plans with aim of the highest quality plan selection.

Results of this study showed that all 4π-NC plans had better performance than VMAT. Although superiority of 4π-NC technique over VMAT is not a surprising result, automated 4π-NC plan had better PTV coverage and higher plan quality metric than manual 4π-NC plan. Lander et al did a study about the number of patients and KBP performance, since the more training data can be led to better generalization and prediction. They found SVDL is more robust to train with limited patient data [34].

Trajectory optimization of 4π is playing a significant role for treatment planning. Collision between the gantry and couch or patient is a considerable challenge for 4π treatment planning which is called collision zones. Since geometry of patients are different, collisional zone varies for different patients. To this end, Northway performed a simulation study about patient-specific collision zones [35]. In this study, full body scans of patients were obtained by optical scanner with aim of patient's CT scan augmentation. Treatment environment was designed by MATLAB programming language on registered images of optical scanner and CT. These information and virtual treatment environment were aligned with a Linac-based treatment. Additionally, a Cranial phantom was used to test this collision detection system. Confusion matrix was utilized in order to evaluate results such that true detection and false detection can be identified.

The accuracy of this system was 98% which was promising. In terms of collision occurred, this system could find collision occurred with surprisingly 99.99% true positive rate. Consequently, this system could be utilized for collision zones identification with least error.

Although 4π-NC technique was applied on IMRT in vast majority of studies, another study applied 4π technique on VMAT. Subramanian et al performed an investigation on the feasibility of multi-isocentric 4π-VMAT (MI-4π-VMAT) to evaluate dose distribution. Therein, 25 patients with head and neck cancer, who were previously treated via VMAT, were studied to compare MI-4π-VMAT and VMAT [36]. Since the head and neck cancer are categorized as complex anatomical site for treatment, OARs sparing has drawn increased attention.

Results of this study showed that OARs doses were significantly decreased using MI-4$\pi$-VMAT. Reduction in average mean dose for bilateral parotids, oral cavity, pharyngeal constrictors, larynx and upper esophagus were 3, 5, 4.3, 4 and 3.3 Gy, respectively. Furthermore, a significant dose reduction by 6 Gy was observed for brainstem. As a result, this study reported MI-4$\pi$-VMAT has a better performance than VMAT in terms of OARs sparing.

**Is 4$\pi$ technique really Covers 4$\pi$ Solid Angle?**

Although 4$\pi$-NC planning receiving much attention, Sarker et al questioned the feasibility of the 4$\pi$ radiotherapy using C-arm linear accelerators [37]. Total solid angle in Euclidian space which can be covered is 4$\pi$ at the center of a sphere in spherical coordinate system with $0 < \theta < \pi$ and $0 < \Phi < 2\pi$. Geometry of this problem can be supposed as a Cantilever for C-arm Linac. As a result, gantry can rotate in range [-$\pi$, 0, $\pi$] and rotation range of table is ($-\frac{\pi}{2}, 0, \frac{\pi}{2}$). Therefore, total solid angle can be calculated for C-arm Linac as follows:

$$\frac{\int_{\theta=0}^{\pi} \int_{\Phi=0}^{\pi} r^2 \sin\theta \, d\Phi \, d\theta}{r^2} = 2\pi \tag{12}$$

Consequently, Sarker et al are being of this opinion that achieving more than 2$\pi$ solid angle with present design of Linac is not possible and "4$\pi$ radiotherapy" is a misnomer.

## Discussion and Suggestions

The solution of optimization problem of 4$\pi$ can be enhanced by adding more robust penalty term to cost function to obtain more sparser solution. Therefore, new mixed matrix norm such as $\|.\|_{2,1/2}$ and $\|.\|_{2-1}$ can be added to enhance sparsity of solution, since these matrixed norms are more sparser that group Lasso $\|.\|_{2,1}$ [38]. Beam selection in 4$\pi$ problem can be modeled as Travelling salesman problem (TSP), since there is a trajectory that must be obtained which is most optimum. There are 1162 beams with 6º angle between two adjacent beams based on the studies where a small number of them should be selected. The optimum beams are the optimum paths which are selected by salesman. Genetic algorithm [39], particle swarm optimization (PSO) [40]and ant colon optimization (ACO) [41] were successful optimization techniques based on evolutionary theory and swarm intelligence to solve TPS problem.

In conclusion, all studies, which are surveyed, reported 4π-NC technique could have better performance compare with VMAT and conventional IMRT in terms of dose escalation for tumors and OAR sparing. Nevertheless, this method is not pervasive technique in radiation therapy and there few studies which applied this technique clinically.